\documentclass[12pt,preprint]{aastex}

\shorttitle{Arp 220 H53$\alpha$ Radio Recombination Line}
\shortauthors{Rodr\'{\i}guez-Rico et al.}

\begin{document}

\title{VLA H53$\alpha$ radio recombination line observations of the ultraluminous infrared galaxy Arp~220}

\author{C. A. Rodr\'{\i}guez-Rico \altaffilmark{1}}
\email{crodrigu@nrao.edu}

\author{W. M. Goss\altaffilmark{2}}
\email{mgoss@nrao.edu}

\author{F. Viallefond\altaffilmark{3}}
\email{fviallef@maat.obspm.fr}

\author{J.-H. Zhao\altaffilmark{4}}
\email{jzhao@cfa.harvard.edu}

\author{Y. G\'omez\altaffilmark{1}}
\email{y.gomez@astrosmo.unam.mx}

\author{K. R. Anantharamaiah\altaffilmark{5}}

\altaffiltext{1}{Centro de Radioastronom\'{\i}a y Astrof\'{\i}sica, UNAM, Campus 
Morelia, Apdo. Postal 3-72, Morelia, Michoac\'an 58089, 
M\'exico.}

\altaffiltext{2}{National Radio Astronomy Observatory, 
Socorro, NM 87801}

\altaffiltext{3}{LERMA, Observatoire de Paris, 61 Av. de 
l'Observatoire F-75014 Paris}

\altaffiltext{4}{Harvard-Smithsonian Center for Astrophysics, 
60 Garden Street, Cambridge, MA 02138}

\altaffiltext{5}{Raman Research Institute, C.V. Raman Avenue, 
Bangalore, 560 080, India. Deceased 2001, October 29}

\begin{abstract}
We present high angular resolution ($0\rlap.{''}7$)
observations made with the Very Large Array (VLA) 
of the radio recombination line (RRL) H53$\alpha$ 
and radio continuum emission at 43~GHz from the ultraluminous 
infrared galaxy (ULIRG) Arp~220. 
The 43~GHz continuum emission shows a compact structure ($\sim 2''$) 
with two peaks separated by $\sim 1''$,
the East (E) and West (W) components, that correspond to each galactic nucleus 
of the merger.
The spectral indices for both the E and W components,
using radio continuum images at 8.3 and 43~GHz are 
typical of synchrotron emission ($\alpha \sim -1.0$).
Our 43 GHz continuum and H53$\alpha$ line observations
confirm the flux densities predicted by the models proposed by
\citet{An00}. This agreement with the models implies the 
presence of high-density ($\sim 10^5$~cm$^{-3}$) compact HII 
regions ($\sim 0.1$~pc) in Arp~220. 
The integrated H53$\alpha$ line emission is stronger toward 
the non-thermal radio continuum peaks, which are also 
coincident with the peaks of molecular emission of the H$_2$CO.
The coincidence between the integrated H53$\alpha$ and the H$_2$CO maser 
line emission suggests that the recent star forming regions, 
traced by the high density gas, are located
mainly in regions that are close to the two radio continuum peaks.
A velocity gradient of $\sim 0.30$~km~s$^{-1}$~pc$^{-1}$ in the H53$\alpha$ RRL 
is observed toward the E component and a second velocity 
gradient of $\sim 0.15$~km~s$^{-1}$~pc$^{-1}$ is detected toward the W component.
The orientations of these velocity gradients are in agreement 
with previous CO, HI and OH observations. 
The kinematics of the high-density ionized gas traced by the H53$\alpha$ line 
are consistent with two counter rotating disks as suggested by the CO and HI observations.

\end{abstract}


\section{INTRODUCTION}

Arp~220 (IC 4553/4; UGC 9913; IRAS 15327+2340) has an infrared luminosity 
of $\sim 1.5 \times 10^{12}$~L$_{\odot}$, as determined from IRAS $25-100$~$\mu$m 
observations (Soifer et al.~1984), and therefore is classified as an ULIRG.
At a distance of $\sim 70$~Mpc (H$_o= 75$~km~s$^{-1}$~Mpc$^{-1}$), Arp~220 is characterized 
in the optical by tidal tails interpreted as the signature of a merging process 
(Surace, Sanders \& Evans 2000).
The large visual extinction toward Arp 220, which could be
A$_{V} > 100$ mag \citep{Go04},
precludes the observations of the nuclear regions 
at optical and even at infrared frequencies.
Radio frequency observations do not suffer from dust extinction,
providing a powerful tool to investigate the kinematics, morphology 
and physical properties of the gas in the dust-obscured 
regions of Arp 220.
Radio observations reveal a double-nucleus system separated 
by $\sim$1$''$ ($\sim 350$~pc) as observed in CO and HI \citep{Mu01,Sa99}.
The existence of multiple luminous radio supernovae (RSN) 
in the nuclear regions of Arp 220 suggests that it is mainly
starburst-powered \citep{Smi98}.
However, compact OH maser and X-rays emission
suggest the presence of AGN activity in the W nucleus
\citep{No85,Lo98,Cl02}.

High angular resolution ($0\rlap.{''}5$) CO observations carried out by 
\citet{Sa99} suggest that the two nuclei are 
counter-rotating with respect to each other and are embedded in a 
kiloparsec-size gas disk, which rotates around their dynamical center. 
The counter-rotation is also observed in higher angular resolution 
($0\rlap.{''}22$) observations of HI absorption \citep{Mu01},
 which show that the rotation is not coplanar and
proposed a model in which the two nuclei are in the final state of merging. 
Counter-rotation may provide the mechanism to get rid of angular 
momentum, a condition for the two nuclei to merge.
The senses of the velocity gradients, as determined from
OH maser emission observations \citep{Ro03}, are consistent
with CO molecular emission and HI absorption observations \citep{Sa99,Mu01}.

Arp~220 is the most distant system from which radio recombination lines (RRLs)
have been observed \citep{Zh96, An00}. 
VLA observations of the centimeter wavelength 
RRL H92$\alpha$ have been reported by \citet{Zh96} and \citet{An00} 
with angular resolutions of 4$''$ and 1$''$, respectively.
\citet{An00}, using a single dish telescope (IRAM 30m), 
detected the emission of the millimeter wavelength RRLs H42$\alpha$, H40$\alpha$, 
and H31$\alpha$ and with the VLA established upper limits for the RRLs 
H167$\alpha$ and H165$\alpha$.
\citet{An00} constructed a model using three density components of
ionized gas in order to reproduce the observed RRL and radio 
continuum intensities, obtaining a star formation rate (SFR) in 
Arp~220 of $\sim 240$~M$_{\odot}$~yr$^{-1}$.
The existence of a high-density component of ionized gas,
as deduced from these models, has been interpreted as 
evidence for recent star formation in Arp~220. 
In particular, the RRL H53$\alpha$ can be used as a tracer of
thermal high-density gas ($\sim 10^5$~cm$^{-3}$).
The star formation in this system may be
supported by multiple starbursts with a high star formation rate.

In this paper we present VLA observations of Arp~220 in the 
millimeter regime of the RRL H53$\alpha$ (7~mm) as well as 
the radio continuum at 43~GHz with subarcsecond 
($0\rlap.{''}7$) angular resolution.
The observations are described in \S 2.
In Section~3 the results obtained from the 43~GHz radio
continuum and the RRL H53$\alpha$ are presented.
In \S~4 the implications of these results are discussed.
The conclusions are presented in \S~5.

\section{VLA Observations of the H53$\alpha$ line.}

The RRL H53$\alpha$ (43~GHz) was observed with the VLA
in the C configuration on October 13 and 25, 2002. 
The flux density scales were determined using observations 
of J1331+305 (1.49~Jy).
The fast-switching mode was used in order to track
the phase variations induced by the troposphere 
using cycles of 230~s on Arp~220 and 40~s on the phase reference 
source J1540+147 ($\sim 1$~Jy).
The bandpass response of the instrument is frequency dependent 
and observations of a bandpass calibrator were required for 
each frequency window.
This bandpass response was corrected using observations of J1229+020 
($17.5 \pm 0.5$~Jy).
Five frequency windows, each of these with 15 spectral channels 
($\sim 22$~km~s$^{-1}$ each), 
were used to observe the broad RRL.
The total velocity coverage is $\sim$1000~km~s$^{-1}$.
The different frequency windows were centered at 42114.9, 
42135.1, 42164.9, 42185.1 and 42214.9~MHz. 
The on-source integration time was $\sim$2~hrs for each frequency window.
To avoid possible phase decorrelation at 7~mm,
the calibration of the data was performed correcting 
for the phases in a first step and subsequently correcting for both amplitude 
and phase. 
The spectral line data were further calibrated by applying the solutions 
obtained from the self-calibration performed on the continuum channel (which 
contains the average of the central 75\% of the bandpass) of each frequency window.
The spectral line observations of the second day were not combined with those of
the first day since the variable content of water vapor in the troposphere caused 
decorrelation of phases in the second and third frequency windows.
A critical step in the reduction of the data is to determine the 
bandpass shape by normalizing the bandpass using the channel 0 data.
Five line images, one for each frequency window of the first day, 
were produced using a weighting scheme intermediate between natural and uniform 
to obtain a circular beam of $0\rlap.{''}7$.
The five line cubes were regridded in frequency and combined 
into a single line cube in the GIPSY environment.
The method used to combine different windows is described by \citet{Ro04}.
The data were then Hanning-smoothed to minimize the Gibbs effect
and a final velocity resolution of $\sim 44$~km~s$^{-1}$ was achieved.
The parameters of the observations are given in Table 1.
The resulting line cube was further processed using IMLIN from AIPS 
to subtract the continuum emission.
A first order polynomial fit was used for the 
line-free channels, two on the low-velocity edge and five on the
high velocity-edge of the bandpass.
Previous Arp 220 observations of the
RRL H42$\alpha$ \citep{An00}, CO(2-1) emission line \citep{Sa99} 
and HI absorption \citep{Mu01}, show that the full width 
at zero intensity expected for the H53$\alpha$ line is
in the range from $\sim 5000$ to $5700$~km~s$^{-1}$.
The velocity range covered by the H53$\alpha$ line observations 
is $5100 - 6100$~km~s$^{-1}$, consequently loosing the continuum 
that corresponds to the blueshifted side of the line.
We performed a test using the line-free channels on the blueshifted side of
the line as a baseline. 
We have determined that if weak H53$\alpha$ line emission with intensity $< 3$~mJy
($3 \sigma$) is present in the velocity range of $5000 - 5100$~km~s$^{-1}$
the full width at half maximum (FWHM) is not affected.
The fit for the H53$\alpha$ line is in agreement with the fit
obtained using the continuum for the low velocity edge of the bandpass.
The H53$\alpha$ (43~GHz) data have been compared with
the RRL H92$\alpha$ (8.3~GHz) observations taken by \citet{An00} 
with the VLA in the B array.

\section{RESULTS}

\subsection{Radio continuum.}

\begin{figure}[!ht]
\epsscale{0.6}
\plotone{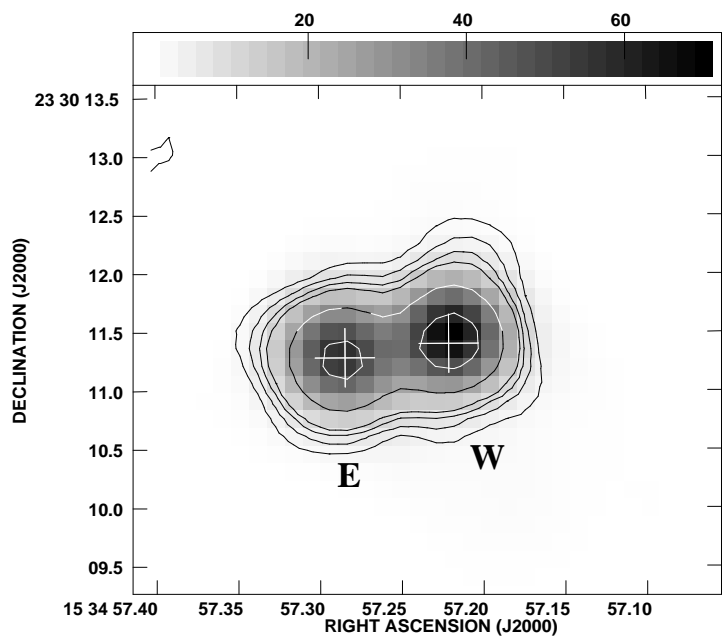}
\caption{Radio continuum image of Arp 220 at 43~GHz (contours) made using VLA observations in the C array
superposed on the continuum image at 8.3~GHz (gray scale) using the VLA in the B array, both images made 
with angular resolution of $0\rlap.{''}7$.
The contour levels (43~GHz) are drawn at -3, 3, 5, 7, 9, 18, and 36 times the rms noise 0.3~mJy~beam$^{-1}$.
The gray scale (8.3~GHz) ranges from 0.15 to 70~mJy~beam$^{-1}$. The crosses mark the peak positions 
(given in Table~2) of the 43 GHz continuum.
The size of the crosses are $0\rlap.{''}5$, which is about five times the error position of the radio continuum peaks.}
\label{cont}
\end{figure}

Figure~\ref{cont} shows the continuum image of Arp~220 at 43~GHz superposed on 
the continuum image at 8.3~GHz, both images  with an angular resolution of $0\rlap.{''}7$.
In this image, the two continuum peaks (E and W)
are clearly resolved at both frequencies.
Using the task IMFIT in AIPS we estimated the properties 
of these two components. The physical parameters at 8.3 and 43 GHz were obtained integrating
over equivalent areas in order to estimate the spectral index $\alpha$ 
(S$_{\nu} \propto \nu^{\alpha}$) of each component.

Table~2 lists the physical parameters of the continuum components E and W
determined at 43~GHz.
The total measured 43~GHz continuum flux density of Arp~220 
is $44 \pm 4$~mJy. The E and W components have continuum 
flux densities of $\sim 17$ and $\sim 21$~mJy, respectively.
There is also a contribution of $\sim 6$~mJy to the total 
flux density from extended emission that was determined
using the task IRING in the AIPS environment 
by integrating the flux density over an angular 
scale of $3''$ to $8''$.
This total continuum flux density at 43~GHz is in agreement
with the value obtained by interpolating from previous observations 
at 32~GHz and 97~GHz (Downes \& Solomon 1998, Anantharamaiah et al.~2000).
The continuum physical parameters at 8.3~GHz of both the E and W 
components in Arp 220 are listed in Table~3.
The positions of each continuum peak are coincident
at 8.3 and 43 GHz.
The two radio nuclei of Arp~220 are separated by $0\rlap.{''}9 \pm 0\rlap.{''}1$ 
at a position angle of 95$^\circ$ (E with respect to N).
The spectral index between these two frequencies for the E component 
is $ -0.96 \pm 0.08$ and for the W component is $ -0.97 \pm 0.07$.
These spectral indices are typical of non-thermal emission.

\subsection{Radio Recombination Line H53$\alpha$}

\begin{figure}[!ht]
\epsscale{1.0}
\plotone{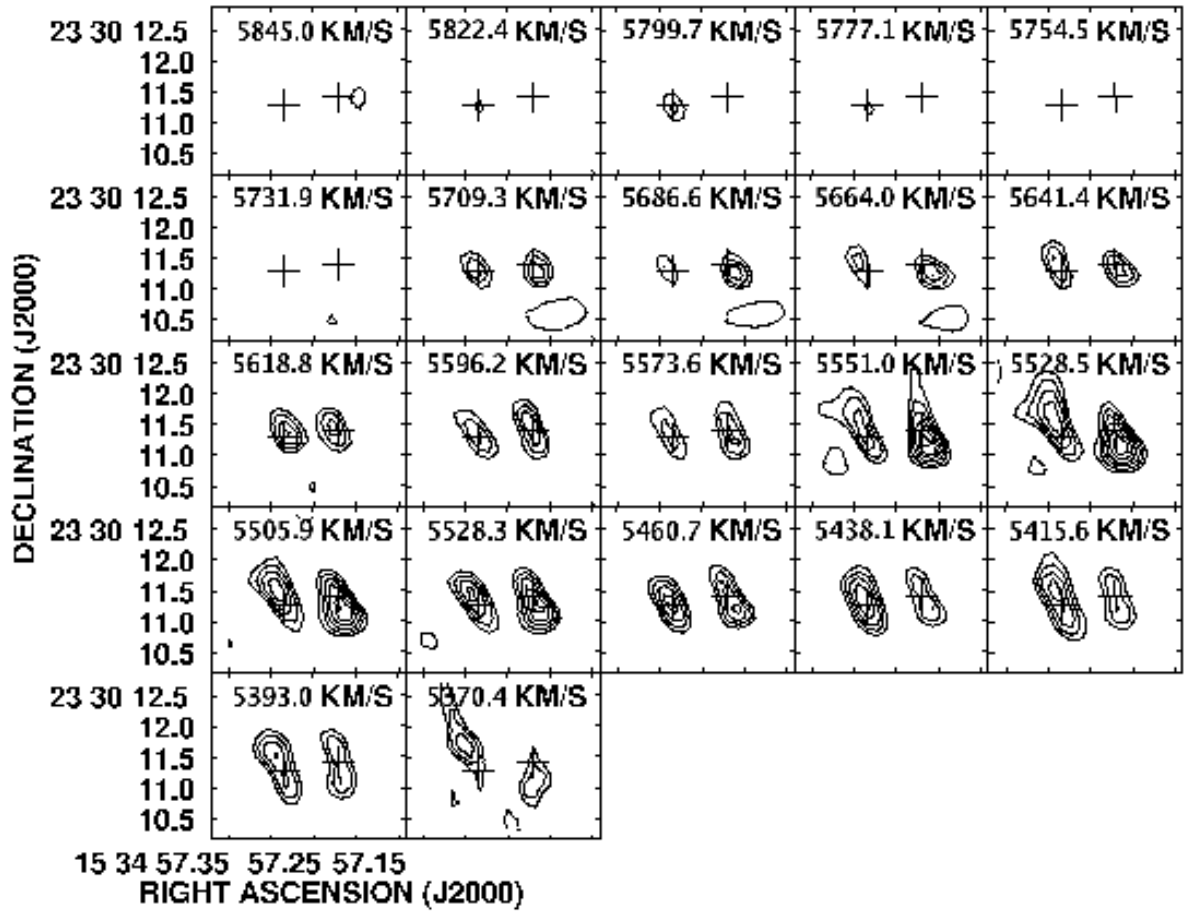}
\caption{H53$\alpha$ continuum subtracted channel images from Arp~220 labeled
according to the central heliocentric velocity of each channel.
Contours are -3, 3, 4, 5, 6, 7, 8, 9, 10 and 11 times 1~mJy~beam$^{-1}$, the rms noise.
The crosses show the position of the two 43~GHz continuum peaks.
The synthesized beam is $0\rlap.{''}7$ FWHM. }
\label{chmap}
\end{figure}

We detected H53$\alpha$ RRL emission ($>3$~mJy) toward both components, 
E and W, of Arp~220 in the velocity range 5370 to 5845~km~s$^{-1}$.
The velocity range over which the H53$\alpha$ line emission is observed
agrees with observations of the RRLs H40$\alpha$, H42$\alpha$, 
and H92$\alpha$ \citep{An00} and the CO(2-1) line \citep{Sa99}.
Figure~\ref{chmap} shows the VLA spectral channel images of H53$\alpha$ 
line emission observed over the central $\sim 500$~pc of Arp~220.
Figure~\ref{mom0} shows the velocity-integrated line emission (moment 0) 
of the H53$\alpha$ line superposed on the moment 0 of the H92$\alpha$ line.
The peaks in the moment 0 of the H53$\alpha$ line emission  
are nearly coincident ($< 0\rlap.{''}5$) with the two radio continuum nuclei.
The moment 0 of the H92$\alpha$ line emission (shown in gray scale) exhibits
two peaks that are close to the continuum peaks and are coincident with
the H53$\alpha$ peaks within $0\rlap.{''}2$.

The total integrated spectrum over the region with 
detectable H53$\alpha$ line emission is shown in Figure~\ref{spec} 
along with the E and W spectra obtained by integrating 
over each of these components.
A single Gaussian fit was carried out for each line profile.
The results from the H53$\alpha$ line fits are listed in Table~4. 
The peak line flux density in column (2),
The FWHM of the line is given in column (3),  
the heliocentric velocity of the line center in column (4) 
and the integrated line emission in column (5).
Based on the fit parameters for the E and W components,
the peak H53$\alpha$ line flux density for the W component
is approximately 30\% stronger than the E component.
The H53$\alpha$ line widths and central velocities are similar for the 
E and W components. The H53$\alpha$ and H92$\alpha$ line widths 
and central velocities determined for the total integrated line emission  
over Arp 220 are in good agreement with each other.
Figure~\ref{mod} shows the results from radio continuum and 
RRL emission models taken from \citet{An00}.
The measurements for the total integrated 43~GHz continuum and H53$\alpha$ 
line emission are shown in Figure~\ref{mod}a and \ref{mod}b, respectively.

Figure~\ref{mom1} shows the H53$\alpha$ line velocity field (moment 1) 
for the ionized gas in Arp 220.
Based on this H53$\alpha$ velocity field, two velocity gradients are 
observed toward the E and W components. The velocity gradient on the E component
is $\sim 0.30 \pm 0.10$~km~s$^{-1}$~pc$^{-1}$, oriented along a P.A. of $\sim 30^{\circ}$. 
The velocity gradient on the W component has been 
marginally detected ($\sim 0.15 \pm 0.10$~km~s$^{-1}$~pc$^{-1}$)
along a P.A. of $\sim 260^{\circ}$.
Due to the lower velocity resolution of the H92$\alpha$ 
line ($230$~km~s$^{-1}$), compared to the velocity resolution 
of the H53$\alpha$ line ($44$~km~s$^{-1}$), a detailed comparison 
of the H53$\alpha$ and the H92$\alpha$ velocity fields is not possible.

\section{DISCUSSION}

The measured separation between the two continuum peaks 
($0\rlap.{''}9 \pm 0\rlap.{''}1 \simeq 320$~pc)
and the P.A. of their separation (95$^{\circ}$) at 43~GHz 
are in good agreement with previous measurements ($\sim 350$~pc) 
at 5, 15 and 23~GHz \citep{No88} and 4.8~GHz \citep{Ba95}. 
The peak positions at 8.3 and 43~GHz agree at the level of $0\rlap.{''}1$.
The spectral index for Arp 220, determined using 
the radio continuum flux density of $\sim 170$~mJy  measured at 
8.3~GHz and the measurement at 43 GHz of $\sim 44$~mJy, is 
$\alpha \sim -1.0$.
On the other hand, the spectral index determined using 
the radio continuum flux density of $\sim 49$~mJy  measured at 
32 GHz (Baan, Gusten \& Haschick 1986) and the measurement 
at 43 GHz (this work) of $\sim 44$~mJy, is $\alpha \simeq 0.36$ ($\pm 0.05$).
\citet{An00} estimated that the thermal radio continuum flux density at 5 GHz
is $\sim 32$~mJy. Using a spectral index value $\alpha=-0.1$ for
the thermal free-free emission the expected thermal flux density at 43~GHz
is $\sim 25$~mJy, which is $\sim 60\%$ of the total measured 
continuum flux density at 43 GHz (see Figure~5).
The flattening of the spectral index, a value of $\alpha$ 
closer to zero when the 32 and 43 GHz measurements are used,
indicates that  the ratio between the thermal emission 
and non-thermal emission is larger at 43~GHz than
at 8.3~GHz.

\citet{An00} modeled the emission of radio continuum and RRL observations 
at different frequencies. In these models \citet{An00} used
three thermally ionized gas components (A1, B1 and D) and non-thermal emission.
Components A1 and D are characterized by
electron densities of $10^3$~cm$^{-3}$, while component B1 has a larger 
electron density of $2.5 \times 10^5$~cm$^{-3}$.
All parameters for each of these components are listed in Table~9
of \citet{An00}.
Figure~\ref{mod} shows the model results from \citet{An00} including the
measurements of the H53$\alpha$ line ($\sim 510 \times 10^{-23}$~W~m$^{-2}$ or 3.7~Jy~km~s$^{-1}$) 
and the continuum flux density at 43 GHz ($\sim 44$~mJy).
The H53$\alpha$ line and 43 GHz continuum flux densities 
are in agreement  with the values predicted in the 
\citet{An00} models, confirming the presence 
of a high-density ionized gas component ($\sim 10^{5}$~cm$^{-3}$) 
composed of compact HII regions with $\sim 0.1$~pc diameter.
According to these models the contribution from the 
dense gas to the H92$\alpha$ line emission is negligible 
and the contribution from the lower density gas to the 
H53$\alpha$ line is $<10\%$. 

\begin{figure}[!ht]
\plotone{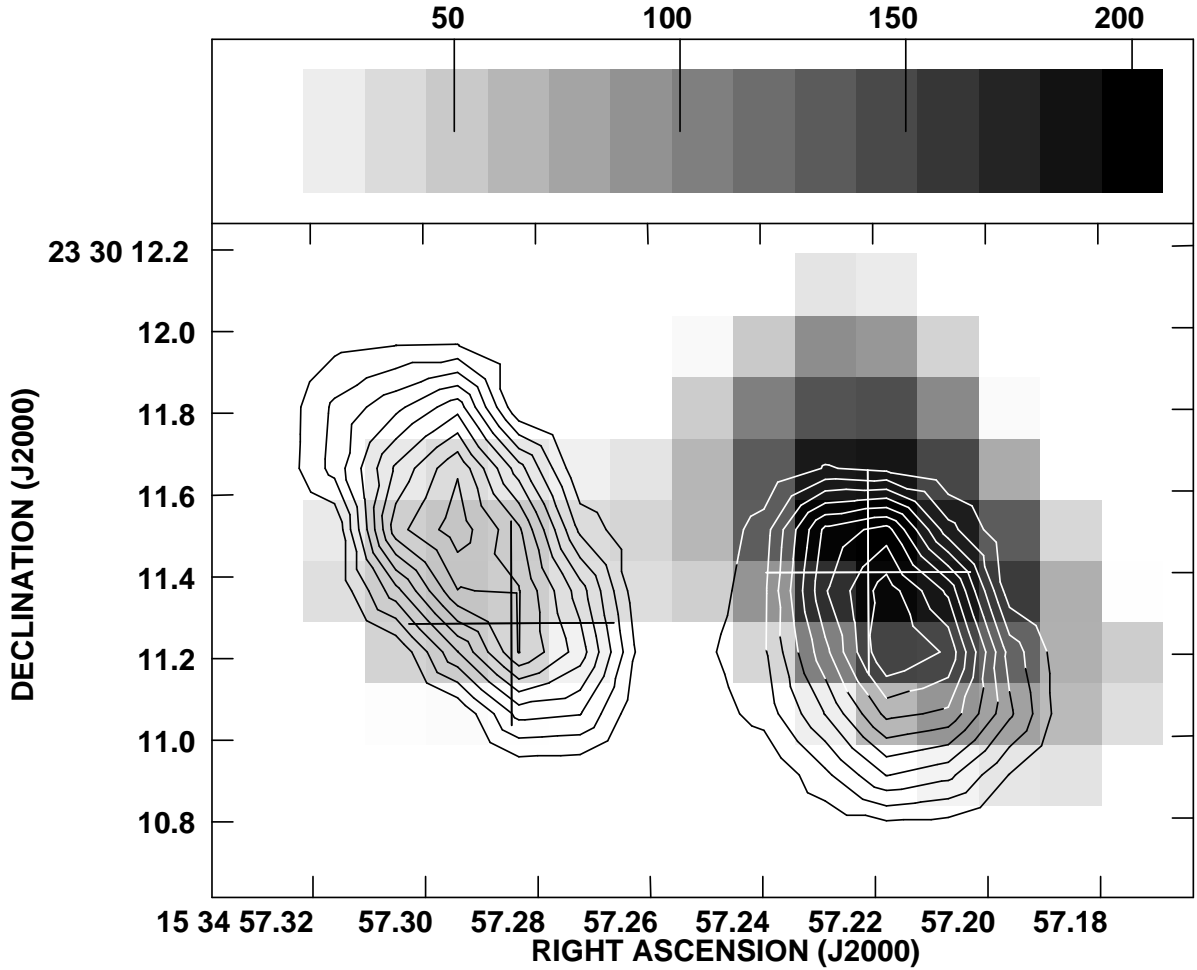}
\caption{Integrated H53$\alpha$ line emission (moment 0) from Arp 220 in
contours with integrated H92$\alpha$ line emission (moment 0) superposed in gray scale.
Contour levels are 10, 30, 50, 70, and 90\% of the peak (1.3~Jy~beam$^{-1}$~km~s$^{-1}$).
The gray scale covers the range $0-200$~Jy~beam$^{-1}$~km~s$^{-1}$.
The crosses show the position of the two 43~GHz continuum peaks.
The synthesized beam for both RRLs is $0\rlap.{''}7$ FWHM. }
\label{mom0}
\end{figure}

\begin{figure}[!ht]
\epsscale{0.5}
\plotone{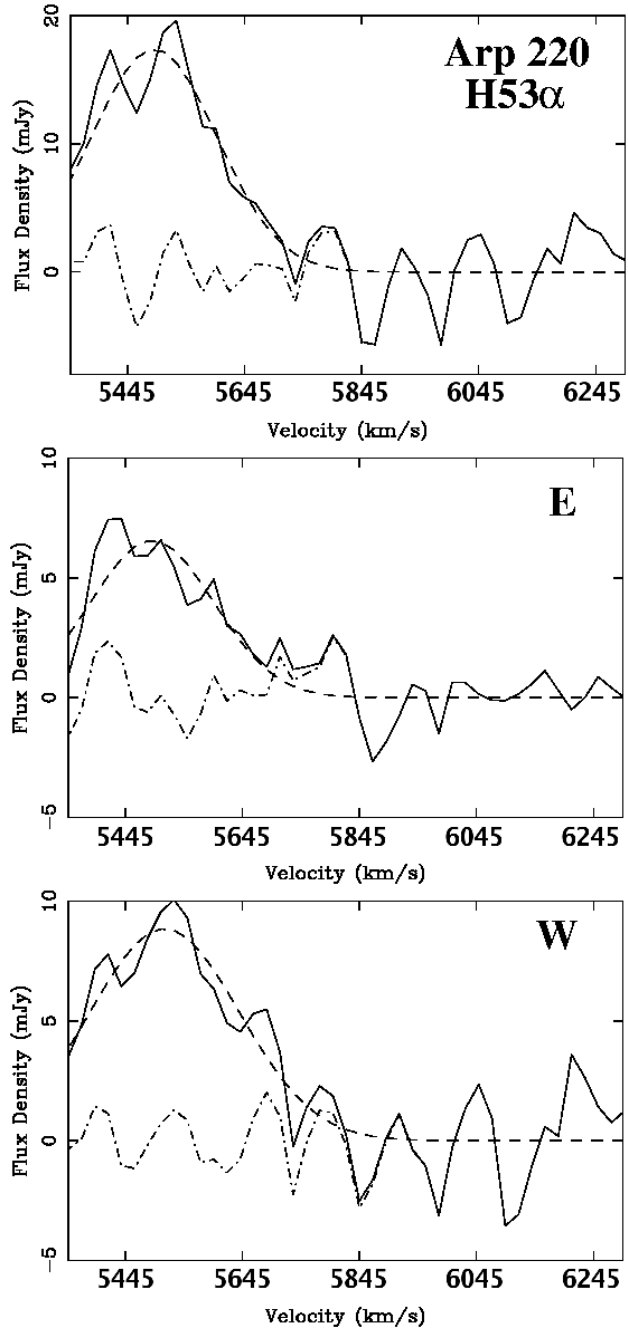}
\caption{H53$\alpha$ recombination line profiles in Arp~220 observed 
using the VLA integrated over the region shown in contours in Figure~1 (top),
over the E component (middle) and over the W component (bottom).
The dashed line shows the resulting Gaussian fit
to the data and the dashed-dotted line indicates 
the residuals from the Gaussian fit.}
\label{spec}
\end{figure}

\begin{figure}[!ht]
\epsscale{0.5}
\plotone{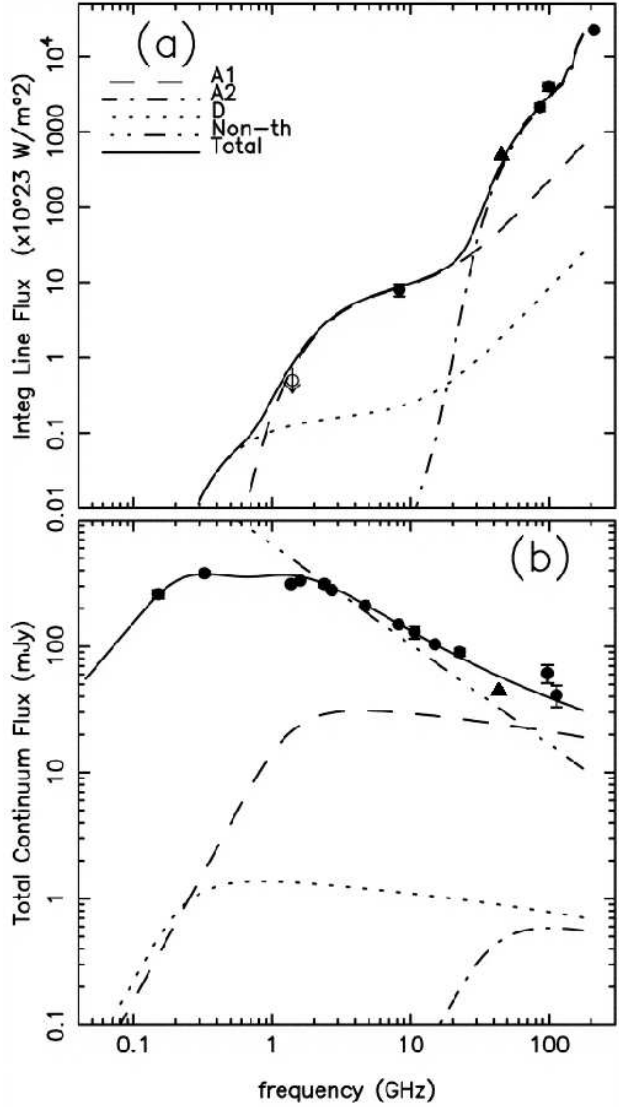}
\caption{Continuum and RRL model results for three ionized gas components 
(A1, A2, and D) in Arp 220 by Anantharamaiah et al.~(2000). 
The recombination line and continuum flux densities are shown in (a)
and (b), respectively.
The contribution from the different components are shown and the 
parameters for each component are given in Table 9 of
Anantharamaiah et al.~(2000).
The observed line data points are listed in Table~2 and 4 of 
Anantharamaiah et al.~(2000).
The filled triangles indicate the velocity integrated H53$\alpha$ 
line flux density in (a) and the 43~GHz continuum flux density in (b)
obtained in the current observations. The size of the triangles represent the
$1 \sigma$ error for the measured values.}
\label{mod}
\end{figure}

\begin{figure}[!ht]
\epsscale{0.7}
\plotone{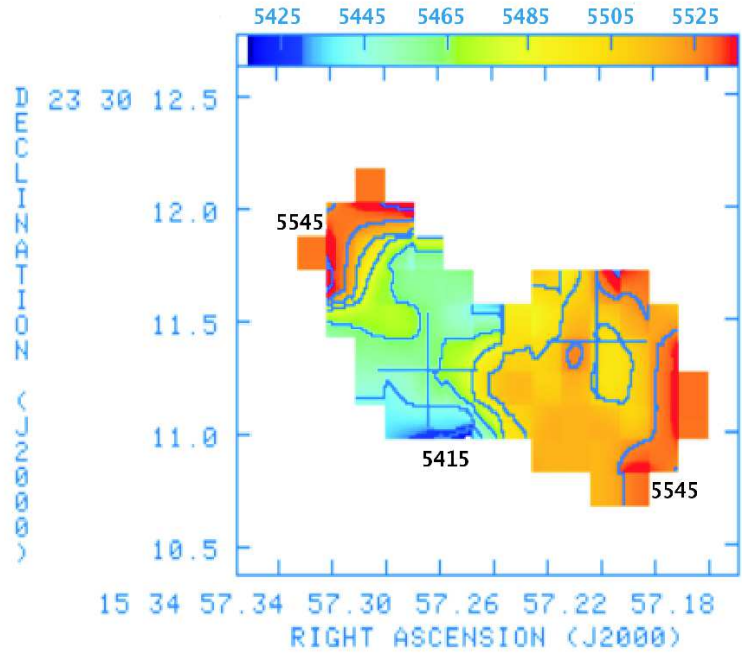}
\caption{Velocity field of Arp~220 as observed in the RRL H53$\alpha$ with the VLA.
The color scale shows the velocity field in the range from 5415 to 
5615~km~s$^{-1}$. Contour levels are 5425, 5445, ..., 5605~km~s$^{-1}$ in steps of 20~km~s$^{-1}$.
The crosses show the position of the two 43~GHz continuum peaks.
The synthesized beam is $0\rlap.{''}7$ FWHM. }
\label{mom1}
\end{figure}

On the other hand, the models of \citet{An00} indicate
that $\sim 50\%$ of the H53$\alpha$ 
line emission in the high-density ionized gas
is due to internal stimulated line emission.
The spatial distribution of the H53$\alpha$ and the H92$\alpha$ lines
is similar toward the E component as can be seen in Figure~\ref{mom0}.
However, a remarkable difference is observed toward 
the W component where the H92$\alpha$ line emission is 
more extended than the H53$\alpha$ line emission.
Near the continuum peaks, the spatial distribution of the 
H53$\alpha$ and H92$\alpha$ RRLs emission suggests that 
the recent star formation is taking place mostly in two regions.
These two regions of line emission are slightly displaced from the respective continuum
peaks: the W line peak is $0\rlap.{''}2 \pm 0\rlap.{''}1$ to the
S of its associated continuum peak and the E line peak
is $0\rlap.{''}3 \pm 0\rlap.{''}1$ to the N of its associated continuum peak.
According to the integrated H53$\alpha$ line emission (see Figure~\ref{mom0}),
the bulk of high-density ionized gas is concentrated in these two regions.
This result is supported by the spatial correlation between the
two velocity-integrated H53$\alpha$ line emission peaks and 
the peaks of the formaldehyde (H$_2$CO) molecular emission \citep{Ba95}.

The models of \citet{An00} derived a SFR in Arp~220 of 
240~M$_{\odot}$~yr$^{-1}$, assuming a mass range of 
$1-100$~M$_{\odot}$ in the Miller-Scalo initial mass function (IMF). 
According to these models, the SFR could 
be as low as 90~M$_{\odot}$~yr$^{-1}$ if the upper mass 
limit in the IMF is reduced to 60~M$_{\odot}$ and the 
Salpeter IMF is used.
Given that the radio continuum and the velocity-integrated line flux density 
are in agreement with the values expected from the models of \citet{An00}, 
the value for the SFR could be in the range of 90 to $240$~M$_\odot$~yr$^{-1}$
and the mass of high-density ionized gas ($\sim 10^5$~cm$^{-3}$) is 
between $ 10^{3}$ and $10^4$~M$_{\odot}$.  

The orientations of the H53$\alpha$ velocity gradients in Arp 220, 
with P.A.$\sim 30^{\circ}$ in the E component and 
P.A.$\sim 260^{\circ}$ in the W component
agree with the CO and HI observations \citep{Sa99,Mu01}.
Based on the P.A. of these two velocity gradients, the CO(2-1) 
line observations were interpreted by \citet{Sa99} as evidence 
of two counter-rotating disks in Arp~220.
An alternative model that consists of a warped gas 
disk that resulted from a merger of two spiral galaxies was also proposed by \citet{Ec01}.
The angular resolution achieved in the H53$\alpha$ line observations ($0\rlap.{''}7$)
is insufficient to discern between these two models.
The velocity structure of the ionized gas based on the 
H53$\alpha$ data show two velocity gradients in the two components of Arp 220.
On the E component the H53$\alpha$ velocity gradient is $\sim 0.30$~km~s$^{-1}$~pc$^{-1}$.
The HI velocity gradient on the E component is $1.01 \pm 0.02$~km~s$^{-1}$~pc$^{-1}$  
\citep{Mu01} and the OH velocity gradient is $0.32 \pm 0.03$~km~s$^{-1}$~pc$^{-1}$ \citep{Ro03}.

The velocity gradient on the E component implies a virial mass 
of $\sim 8 \times 10^{7}~(sin^{-2}i)$~M$_{\odot}$ in a disk of 
radius $\sim$180~pc ($i$ is the inclination).
The total mass calculated from OH observations, for a disk of radius $\sim 80$~pc,
is $\sim 10^7$~M$_{\odot}$ \citep{Ro03}. The value derived from the OH observations
is consitent with the H53$\alpha$ line observations when the different sizes are taken into account.
Based on HI and CO \citep{Sa99,Mu01} observations, the estimates
of the total mass are $\sim 10^9$~M$_{\odot}$.
The total mass estimated using CO and HI observations
is two orders of magnitude larger than the total mass 
calculated from these H53$\alpha$ line observations.
The smaller estimates obtained from the H53$\alpha$ line and OH observations
are explained if the molecular CO and neutral 
HI gas are distributed over a larger region compared to the ionized gas.
In the direction of the W component, the H53$\alpha$ velocity gradient is 
a factor of 10 less than observed in HI absorption and OH maser emission 
\citep{Mu01,Ro03}. However, the orientation and sense of the 
velocity gradient in H53$\alpha$ are consistent with those 
determined from the CO, HI and OH observations \citep{Sa99, Mu01, Ro03}.

\section{CONCLUSIONS.}

We have observed the RRL H53$\alpha$ and radio continuum at 43~GHz 
toward the ULIRG Arp~220 with high angular resolution 
($0\rlap.{''}7$) using the VLA. 
The 43 GHz radio continuum and the RRL H53$\alpha$ 
have been compared with observations of the 8.3 GHz radio continuum and  
H92$\alpha$ line \citep{An00}.

\begin{itemize}
\item  The total 43 GHz radio continuum flux density of Arp 220 is $44 \pm 4$~mJy.
The morphological characteristics observed in the radio continuum at 43 GHz 
agree with previous radio observations made in the range from $1.4$ to $23$~GHz.
In the radio continuum at 43~GHz, Arp 220 exhibits a double-nucleus system that has 
been clearly resolved with a separation of the two radio continuum peaks 
of $\sim 1''$ at P.A. of 95$^{\circ}$.

\item We have determined the spectral indices for both the E and W components,
using radio continuum flux densities at 8.3 and 43~GHz. Both components
have spectral indices typical of synchrotron emission ($\alpha \sim -0.9$), as
expected from extrapolation from the centimeter wavelength range.

\item Based on 43 GHz VLA data, we confirm the predictions of 
the models made by \citet{An00} for RRLs.
The integrated H53$\alpha$ line flux density is 
about a factor of $\sim 50$ times larger than the 
integrated H92$\alpha$ line flux density, in 
agreement with the predictions of \citet{An00}.
Thus, the H53$\alpha$ line traces the high-density 
($\sim 10^5$~cm$^{-3}$) compact HII regions ($\sim 0.1$~pc) in 
Arp~220. 

\item The spatial distributions of the H53$\alpha$ and the 
H92$\alpha$ line emission are similar in the direction of the E component.
On the W component the low density ($\sim 10^3$~cm$^{-3}$) 
ionized gas component is more extended than the high density
($\sim 10^5$~cm$^{-3}$) component.

\item The kinematic and spatial distribution behavior as observed in the 
RRL H53$\alpha$ is in agreement with results reported from 
CO and HI observations \citep{Sa99,Mu01}, supporting the counter-rotation 
of two disks in Arp 220 at smaller scales.
\end{itemize}

The original version was published in the Astrophysical Journal, volume 633, page 198 in year 2005. 
This new version is submitted with the corrected velocity scale.
In Figures 2, 4 and 6 the values for radial velocities were incorrect. All velocities were increased 
by 245~km~s$^{-1}$ to correct values to heliocentric values (optical definition). To obtain the velocities 
using radio definition, add 145~km~s$^{-1}$. Thus the central velocities in Table~4 should be 5495~km~s$^{-1}$ 
(optical definition) or 5395~km~s$^{-1}$ (radio definition). This velocity shift was introduced in the frequency 
reprojection process. The velocity definition in Table 4 is the optical heliocentric definition, as is used by
K. R. Anantharamaiah et al. (ApJ, 537, 613 [2000]). All other quantities in the paper remain unchanged, e.g. 
the radio recombination line and radio continuum integrated properties, as well as distributions.

The National Radio Astronomy Observatory is a facility of the National Science Foundation operated under
cooperative agreement by Associated Universities, Inc.
CR and YG acknowledge the support from UNAM and CONACyT, M\'exico.
The authors thank the referee for helpful comments.

\clearpage

\begin{deluxetable}{cc}
\tablecolumns{2}
\tablewidth{0pc}
\tablecaption{O{\footnotesize BSERVATIONAL }
P{\footnotesize ARAMETERS FOR} Arp~220
}
\tablehead{
\colhead{Parameter} &  \colhead{H53$\alpha$ Line}}
\startdata
Right ascension (J2000)\dotfill                 &       15 34 57.28\\
Declination (J2000)\dotfill                     &       23 30 11.9\\
Angular resolution\dotfill                      &       $0\rlap.{''}7$\\
Total observing duration (hr)\dotfill           &       13\\
Bandwidth (MHz)\dotfill                         &       150\\
Number of spectral channels\dotfill             &       45\\
Center V$_{Hel}$ (km~s$^{-1}$)\dotfill          &       5500\\
Velocity coverage (km~s$^{-1}$)\dotfill         &       1000\\
Velocity resolution (km~s$^{-1}$)\dotfill       &       44\\
Amplitude calibrator\dotfill                    &       J1331+305\\
Phase calibrator\dotfill                        &       J1540+147\\
Bandpass calibrator\dotfill                     &       J1229+020\\
RMS line noise per channel (mJy/beam)\dotfill   &       1\\
RMS, continuum (mJy/beam)\dotfill               &       0.3\\
\enddata
\end{deluxetable}

\clearpage

\begin{deluxetable}{ccccc}
\tablecolumns{6}
\tablewidth{0pc}
\tablecaption{R{\footnotesize ESULTS FROM GAUSSIAN FITTING TO THE 
CONTINUUM EMISSION AT 43~GHZ OF ARP~220}. }
\tablehead{
\colhead{Feature} &  \colhead{RA (J2000)} & \colhead{DEC (J2000)} &  \colhead{Size\tablenotemark{a}, P.A.}  & \colhead{S$_C$ (mJy)} }
\startdata
Arp 220 E \dotfill & $15^{h}34^{m}57\rlap.{^{s}}28 \pm 0\rlap.{^{s}}01$ & 
$23^{\circ}30' 11\rlap.{''}3 \pm 0\rlap.{''}1$ & 
$0\rlap.{''}5 \times 0\rlap.{''}3$, 89$^{\circ}$  & $17 \pm 2$ \\
Arp 220 W \dotfill & $15^{h}34^{m}57\rlap.{^{s}}22 \pm 0\rlap.{^{s}}01$ & 
$23^{\circ} 30' 11\rlap.{''}4 \pm 0\rlap.{''}1$ & 
$0\rlap.{''}4 \times 0\rlap.{''}2$, 114$^{\circ}$ & $21 \pm 2$ \\
\enddata
\tablenotetext{a}{Deconvolved angular size.}
\end{deluxetable}

\clearpage

\begin{deluxetable}{ccccc}
\tablecolumns{6}
\tablewidth{0pc}
\tablecaption{R{\footnotesize ESULTS FROM GAUSSIAN FITTING TO THE 
CONTINUUM EMISSION AT 8.3~GHZ OF ARP~220\tablenotemark{a}}. }
\tablehead{
\colhead{Feature} &  \colhead{RA (J2000)} & \colhead{DEC (J2000)} 
&  \colhead{Size\tablenotemark{b}, P.A.}  & \colhead{S$_C$ (mJy)\tablenotemark{c}} }
\startdata
Arp 220 E \dotfill & $15^{h}34^{m}57\rlap.{^{s}}29 \pm 0\rlap.{^{s}}01$ & 
$23^{\circ}30' 11\rlap.{''}3 \pm 0\rlap.{''}1$ & 
$0\rlap.{''}5 \times 0\rlap.{''}4$, 88$^{\circ}$  & $77 \pm 1$ \\
Arp 220 W \dotfill & $15^{h}34^{m}57\rlap.{^{s}}22 \pm 0\rlap.{^{s}}01$ & 
$23^{\circ} 30' 11\rlap.{''}5 \pm 0\rlap.{''}1$ & 
$0\rlap.{''}5 \times 0\rlap.{''}3$, 107$^{\circ}$ & $89 \pm 1$ \\
\enddata
\tablenotetext{a}{Observations at 8.3~GHz made by \citet{An00}.}
\tablenotetext{b}{Deconvolved angular size.}
\tablenotetext{c}{Continuum flux densities were measured in the same area as the 43~GHz continuum flux densities listed in Table~2.}
\end{deluxetable}

\clearpage

\begin{deluxetable}{ccccc}
\tablecolumns{5}
\tablewidth{0pc}
\tablecaption{R{\footnotesize ESULTS FROM GAUSSIAN FITTING TO THE 
H53$\alpha$ LINE EMISSION OF ARP~220}. }
\tablehead{
\colhead{Feature} & 
\colhead{S$_P$ } & 
\colhead{$\Delta$~V$_{FWHM}$ } &  
\colhead{V$_{Helio}$ }  
&  \colhead{$1.07$~(S$_P$~$\Delta$V$_{FWHM}$}) \\
& (mJy) & (km~s$^{-1}$) & (km~s$^{-1}$) & (W~m$^{-2} \times 10^{-23}$ ) }
\startdata
Arp~220 E   \dotfill &  $6  \pm 1$ & $235 \pm 20$   & $5485 \pm 10$ & $220 \pm 40$\\
Arp 220 W   \dotfill &  $8  \pm 1$ & $265 \pm 30$   & $5515 \pm 10$ & $320 \pm 50$\\
Arp 220 E+W \dotfill &  $16 \pm 2$ & $230 \pm 20$   & $5495 \pm 10$ & $560 \pm 80$\\
\enddata
\end{deluxetable}

\clearpage

\clearpage

\clearpage

\clearpage

\clearpage

\clearpage

\appendix

\end{document}